\documentclass[preprint2]{emulateapj}
\usepackage{apjfonts}
\submitted{Received 2011 February 27; accepted 2010 April 5; published --}
\journalinfo{Accepted to ApJ Letters 04/05/2011}
\newcommand{\FeI}{\ion{Fe}{1}}
\newcommand{\CaII}{\ion{Ca}{2}}

\newcommand{\ms}{m$\,$s$^{-1}$}

\shorttitle{Rapid Enhancement of Sheared Evershed Flow}
\shortauthors{Deng et al.}

\begin{document}
\title{Rapid Enhancement of Sheared Evershed Flow Along the Neutral Line Associated with an X6.5 Flare Observed by Hinode}

\author{Na Deng\altaffilmark{1,2}, Chang Liu\altaffilmark{2}, Debi Prasad Choudhary$^{1}$ and Haimin Wang\altaffilmark{2}}
\affil{1. Physics and Astronomy Department, California State University Northridge, 18111 Nordhoff St., Northridge, CA 91330, USA; na.deng@csun.edu, debiprasad.choudhary@csun.edu}
\affil{2. Space Weather Research Laboratory, New Jersey Institute of Technology, University Heights, Newark, NJ 07102, USA; chang.liu@njit.edu, haimin@flare.njit.edu}

\begin{abstract}
We present G-band and \CaII\ H observations of NOAA AR 10930 obtained by Hinode/SOT on 2006 December 6 covering an X6.5 flare. Local Correlation Tracking (LCT) technique was applied to the foreshortening-corrected G-band image series to acquire horizontal proper motions in this complex $\beta\gamma\delta$ active region. With the continuous high quality, spatial and temporal resolution G-band data, we not only confirm the rapid decay of outer penumbrae and darkening of the central structure near the flaring neutral line, but also unambiguously detect for the first time the enhancement of the sheared Evershed flow (average horizontal flow speed increased from 330$\pm$3.1 to 403$\pm$4.6 \ms) along the neutral line right after the eruptive white-light flare. Post-flare \CaII\ H images indicate that the originally fanning out field lines at the two sides of the neutral line get connected. Since penumbral structure and Evershed flow are closely related to photospheric magnetic inclination or horizontal field strength, we interpret the rapid changes of sunspot structure and surface flow as the result of flare-induced magnetic restructuring down to the photosphere. The magnetic fields turn from fanning out to inward connection causing outer penumbrae decay, meanwhile those near the flaring neutral line become more horizontal leading to stronger Evershed flow there. The inferred enhancement of horizontal magnetic field near the neutral line is consistent with recent magnetic observations and theoretical predictions of flare-invoked photospheric magnetic field change.

\end{abstract}


\section{INTRODUCTION}\label{sec:introduction}

Solar flares are processes of magnetic reconnection accompanied by violent release of accumulated magnetic free energy in the solar atmosphere. Recently, increasing observational evidence has shed new light on the flare associated magnetic field change, which occurs not only in the corona as expected, but also extends down to the dense photosphere where the magnetic fields were assumed anchored and insusceptible to coronal eruption \citep[e.g.,][]{Kopp+Pneuman1976SoPh...50...85K, Choudhary+Gary1999SoPh..188..345C}. Early low cadence vector magnetograph observations have hinted at possible flare-related changes of photospheric transverse field and magnetic shear near the flared neutral lines \citep{WangH1992SoPh..140...85W, Ambastha+etal1993SoPh..148..277A, Wang+etal1994ApJ...424..436W, Chen+etal1994SoPh..154..261C, Hagyard+etal1999SoPh..184..133H, Li+etal2000PASJ...52..483L}. When continuous high cadence line-of-sight (LOS) photospheric magnetograms become available, rapid and permanent magnetic flux changes are found spatially and temporally associated with flare occurrence \citep{Spirock+etal2002ApJ...572.1072S, Wang+etal2002ApJ...576..497W, Yurchyshyn+etal2004ApJ...605..546Y, Sudol+Harvey2005ApJ...635..647S, WangH2006ApJ...649..490W}. The interpretations of those impulsive flux changes in different flare events were inconclusive then. On the other hand, from white-light (WL) observations, rapid outer penumbral decay and central (i.e., near flaring neutral line; throughout this letter, the term ``central'' always refers to near flaring neutral line) structure darkening are found quite commonly in complex $\delta$ sunspots right after major flares \citep{WangH+etal2004ApJ...601L.195W, Deng+etal2005ApJ...623.1195D, LiuC+etal2005ApJ...622..722L, ChenW+etal2007ChJAA...7..733C}. Since sunspots are magnetic in nature and their structure is closely related to the local magnetic field configuration, the authors interpret the sudden change of sunspot structure as the manifestation of magnetic field restructuring at the photosphere. In particular, \citet[][Fig.12]{LiuC+etal2005ApJ...622..722L} present a schematic reconnection picture that explains most of the observational aspects, where the original fanning out field lines of opposite polarities get connected over the neutral line and plunged downward after the flare, resulting in more vertical fields in the outer part and more horizontal fields in the central region compared to pre-flare configuration. Recent precise vector magnetic field measurements endorse the concept of this reconnection picture. For instance, \citet{Li+etal2009ScChG..52.1702L} find that the mean inclination angle (with respect to surface normal) in the outer decayed penumbral regions decreases (i.e., becomes more vertical) while that in the central darkened areas increases (i.e., becomes more inclined) after a major flare. Focusing on the flare-induced changes in horizontal magnetic fields, \citet{WangJ+etal2009ApJ...690..862W} detected substantial strengthening of horizontal magnetic field in the central region and weakening in patches in the outskirts of a $\delta$ sunspot group.

The clearly observed flare-associated photospheric magnetic field changes have also drawn the attention of theorists. Quantitatively assessing the change of Lorentz force that needs to be rebalanced during magnetic eruptions, \citet{Hudson+Fisher+Welsch2008ASPC..383..221H} and \citet{Fisher+etal2010arXiv1006.5247F} conclude that the photospheric magnetic fields near the flaring neutral line should undergo a collapse (i.e., become more horizontal) along with a downward (inward) jerk resulting from drastic coronal magnetic reconfigurations. This is in principle consistent with the reconnection picture presented by \citet{LiuC+etal2005ApJ...622..722L}. \citet{Wang+Liu2010ApJ...716L.195W} synthesized the vector and LOS magnetic field observations covering major flares and find that 23 out of 24 events showed signatures of collapsed or more horizontal magnetic fields near flared neutral lines, which thus strongly supports the aforementioned theoretical prediction. We notice that the enhancement of central horizontal photospheric magnetic fields is also present in the tether-cutting flare model \citep{Moore+etal2001ApJ...552..833M} after the low-lying short loop is formed in the first stage reconnection. The collapse of near surface fields is also visible in the animations of three-dimensional MHD simulations of and right after coronal eruptions \citep[e.g.,][]{FanY2005ApJ...630..543F, FanY2010ApJ...719..728F, Pariat+etal2009ApJ...691...61P}.

Penumbrae are indicators of horizontal magnetic fields in the photosphere. Penumbral fibrils are known to follow the magnetic azimuth direction.
For example, sheared (i.e., tangent to umbrae) penumbrae and associated sheared Evershed flows are often observed along the neutral line of complex sunspots with multiple polarities \citep{yang+etal2004, Deng+etal2006ApJ...644.1278D, Denker+etal2007SoPh..245..219D, Tan+etal2009ApJ...690.1820T}. In addition, the penumbral structure and carried Evershed flow are closely related to the magnitude of magnetic inclination. Recent high-resolution observations \citep[e.g.,][and references therein]{Ichimoto2010mcia.conf..186I} and significant breakthrough made in realistic three-dimensional numerical MHD simulations \citep{Rempel+etal2009Sci...325..171R, Kitiashvili+etal2009ApJ...700L.178K} have revealed the thermal magnetoconvective nature of penumbral Evershed effect in the presence of strong (kilogauss) and inclined (averagely larger than 45$^\circ$) magnetic field. Moreover, the filamentary pattern and the speed of horizontal surface flow of the simulated penumbra are strongly controlled by both strength and inclination of the embedded magnetic field with inclination playing the most important role. In summary, the penumbra and the Evershed flow are stronger with larger magnetic inclination or stronger horizontal fields within a certain range \citep{Hurlburt+etal1996ApJ...457..933H, Hurlburt+etal2000SoPh..192..109H, Rempel+etal2009ApJ...691..640R, Kitiashvili+etal2009ApJ...700L.178K}. Using Hinode Spectro-Polarimeter (SP) and G-band observations, \citet{Deng+etal2011arXiv1102.3164D} statistically examined the relationship between magnetic field structure and penumbral size as well as horizontal Evershed flow speed and find that the latter two are positively correlated with magnetic inclination or horizontal field strength. This observational result is thus consistent with the aforementioned simulations and opens a window for penumbra and Evershed flow to diagnose magnetic configuration in the photosphere. Being aware that the above simulations and observations deal with simple sunspots, we believe that the penumbrae in complex sunspots should share similar underlying physics.

\section{OBSERVATION AND DATA REDUCTION}\label{sec:observation}
The X6.5 flare that occurred in NOAA AR 10930 was a two-ribbon white-light eruptive flare peaked around 18:43~UT on 2006 December 6 and associated with a powerful Moreton wave and coronal mass ejection \citep{Balasubramaniam+etal2010ApJ...723..587B}. It is the largest flare event so far well captured by Solar Optical Telescope \citep[SOT;][]{Tsuneta+etal2008SoPh..249..167T, Suematsu+etal2008SoPh..249..197S, Shimizu+etal2008SoPh..249..221S} of Hinode satellite \citep{Kosugi+etal2007SoPh..243....3K}. Except for a 16-minute data gap (from 18:57 to 19:13~UT) right after the main hard X-ray (HXR) phase of the flare, continuous high resolution ($\sim$ 0.2 \arcsec) G-band (430.5~nm) and \CaII\ H (396.8~nm) broadband filtergrams with 2-minute cadence were taken several hours before until one hour after the flare. A SOT/SP scan of the region with full Stokes spectra of \FeI\ 630.2~nm lines \citep{Ichimoto+etal2008SoPh..249..233I} was also taken about 3 hours before the flare.

\begin{figure}[t]
  \epsscale{1.15}
  \plotone{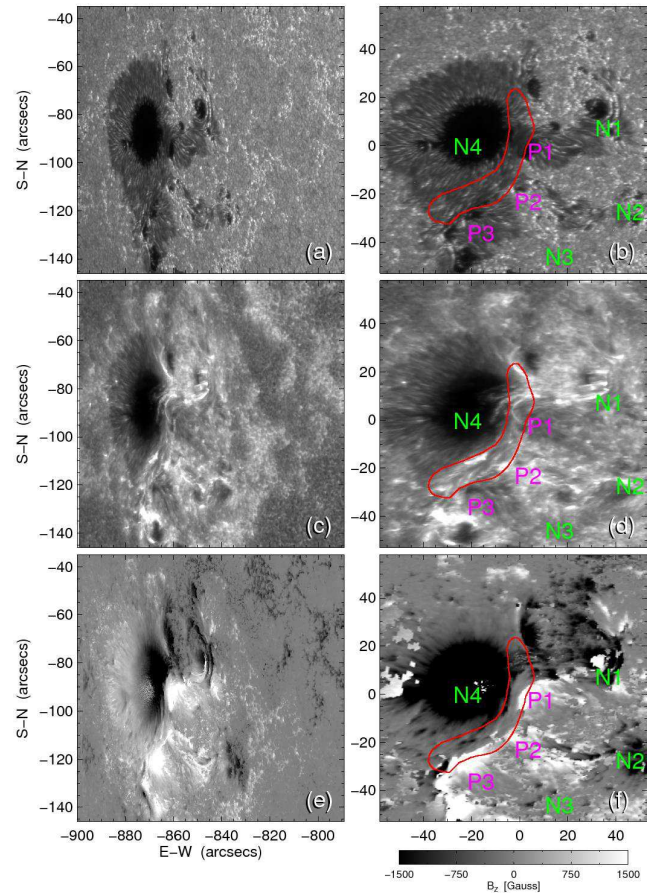}
  \caption{Hinode/SOT observations of NOAA AR 10930 at 15:00~UT on 2006 December 6 in G-band (\textsl{Top row}) and \CaII\ H (\textsl{Middle row}) filtergrams, as well as SP B$_{los}$ and B$_z$ magnetograms (\textsl{Bottom row}). The left and right columns show the original observed and projection-corrected images, respectively. The red contour, which is roughly centered at the flaring neutral line, outlines our region of interest (ROI) where the sheared Evershed flow along the neutral line dramatically enhanced after the X6.5 flare. P1-P3 and N1-N4 denote the positive and negative patches to be discussed for magnetic connection later.}
  \label{FIG:DEPROJ}
\end{figure}

\begin{figure*}[t]
  \epsscale{1.15}
  \plotone{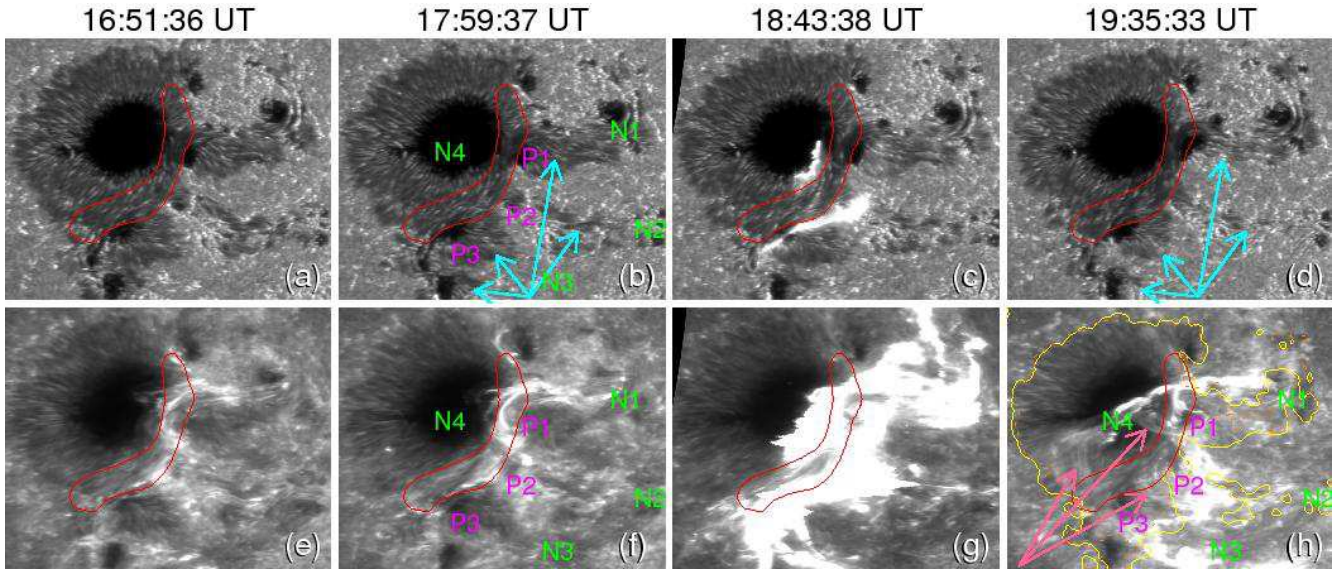}
  \caption{Time evolution of NOAA AR 10930 in G-band (\textsl{Top row}) and \CaII\ H (\textsl{Bottom row}) images. The blue arrows on panels $(b)$ and $(d)$ point out the outer penumbral regions that decayed after the flare. The solid yellow and dotted orange contours on panel $(h)$ outline the G-band sunspot border in pre- and post-flare phases, respectively, to indicate the decayed penumbral regions. The pink arrows point to the newly connected post-flare loops over the neutral line.}
  \label{FIG:IMGS}
\end{figure*}

\begin{figure*}
  \epsscale{1.15}
  \plotone{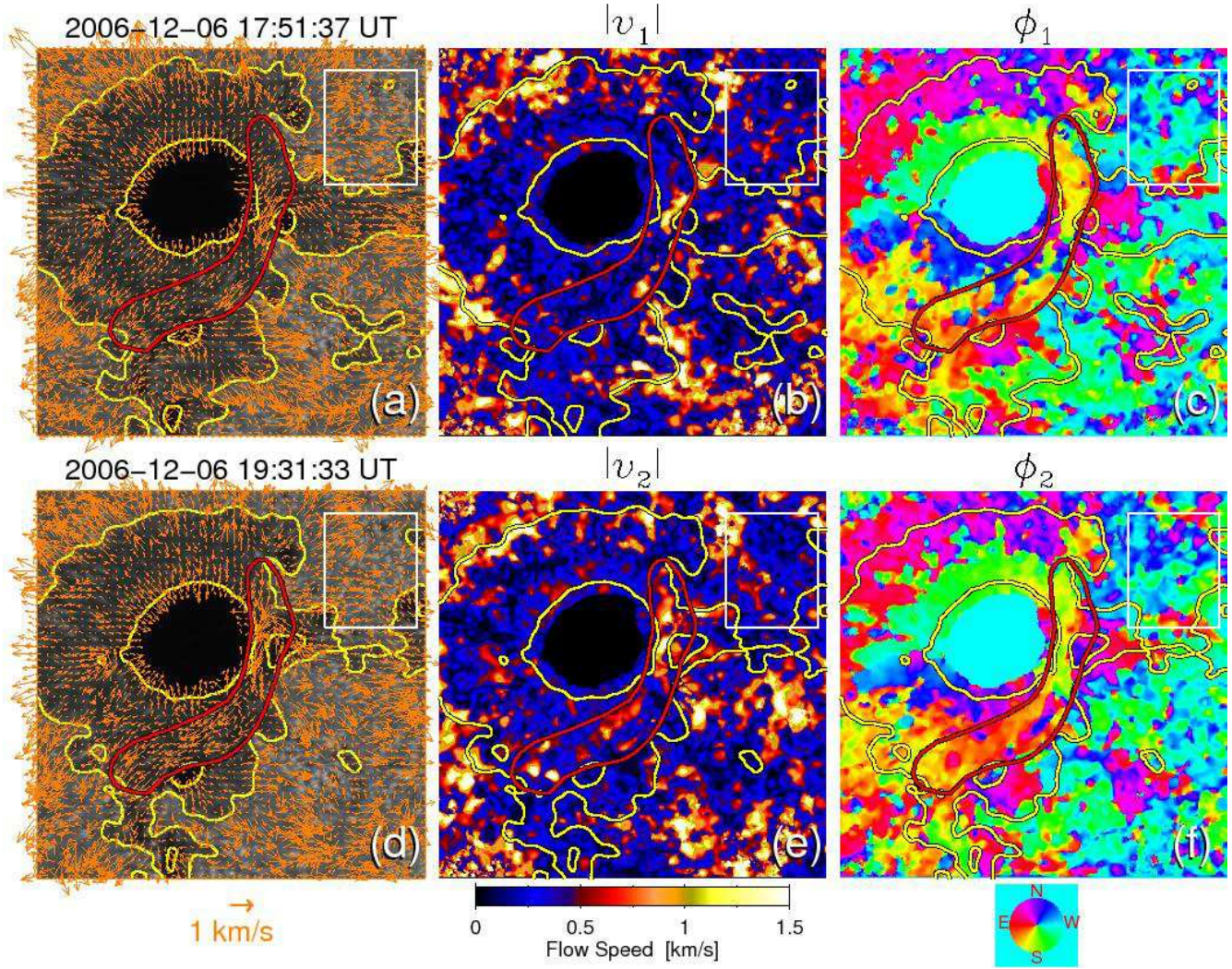}
  \caption{Horizontal proper motions derived from LCT in NOAA AR 10930 at moments $\sim$50 minutes before (\textsl{Top row, footnoted as 1}) and $\sim$50 minutes after (\textsl{Bottom row, footnoted as 2}) the flare peak. The speed ($|\upsilon|$) and azimuth ($\phi$) maps of the horizontal proper motions are shown in the 2nd and 3rd columns. The upper-right white box outlines a reference region to be used in Fig.~\ref{FIG:PLOT}.}
  \label{FIG:VELO}
\end{figure*}

\begin{figure*}
  \epsscale{1.1}
  \plotone{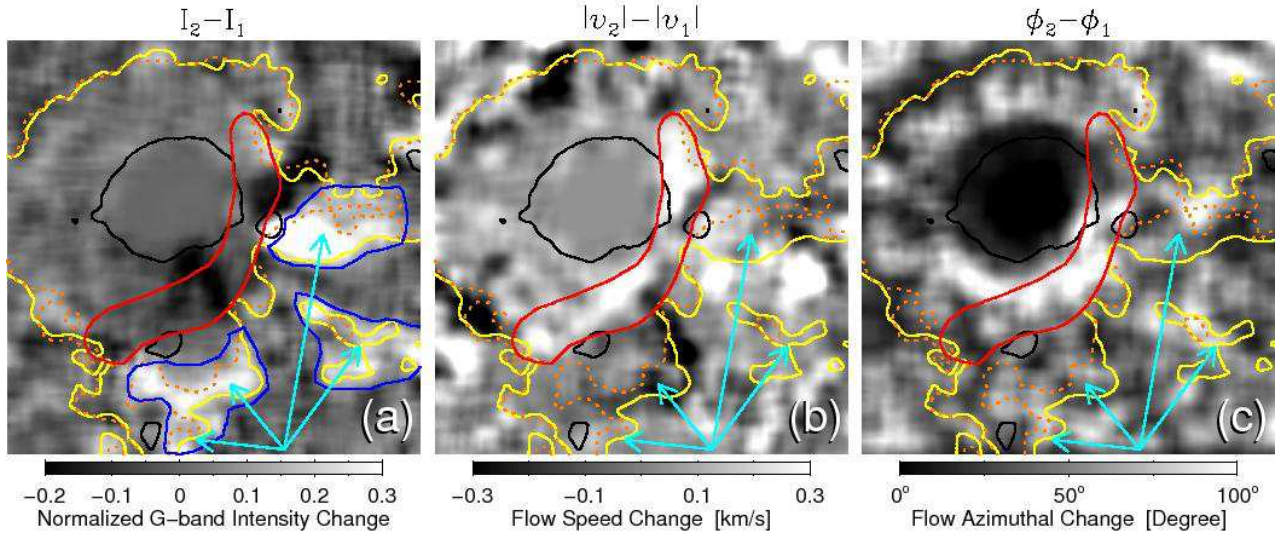}
  \caption{The difference maps between post-(19:31~UT, footnoted as 2) and pre-(17:51~UT, footnoted as 1) flare data showing the flare-associated changes in $(a)$ G-band intensity, $(b)$ horizontal flow speed, and $(c)$ horizontal flow azimuth. These difference maps are results of spatial smooth using a 5.6$^{\prime\prime}\times$5.6$^{\prime\prime}$ window.}
  \label{FIG:DIF}
\end{figure*}

\begin{figure}[t]
  \epsscale{1.2}
  \plotone{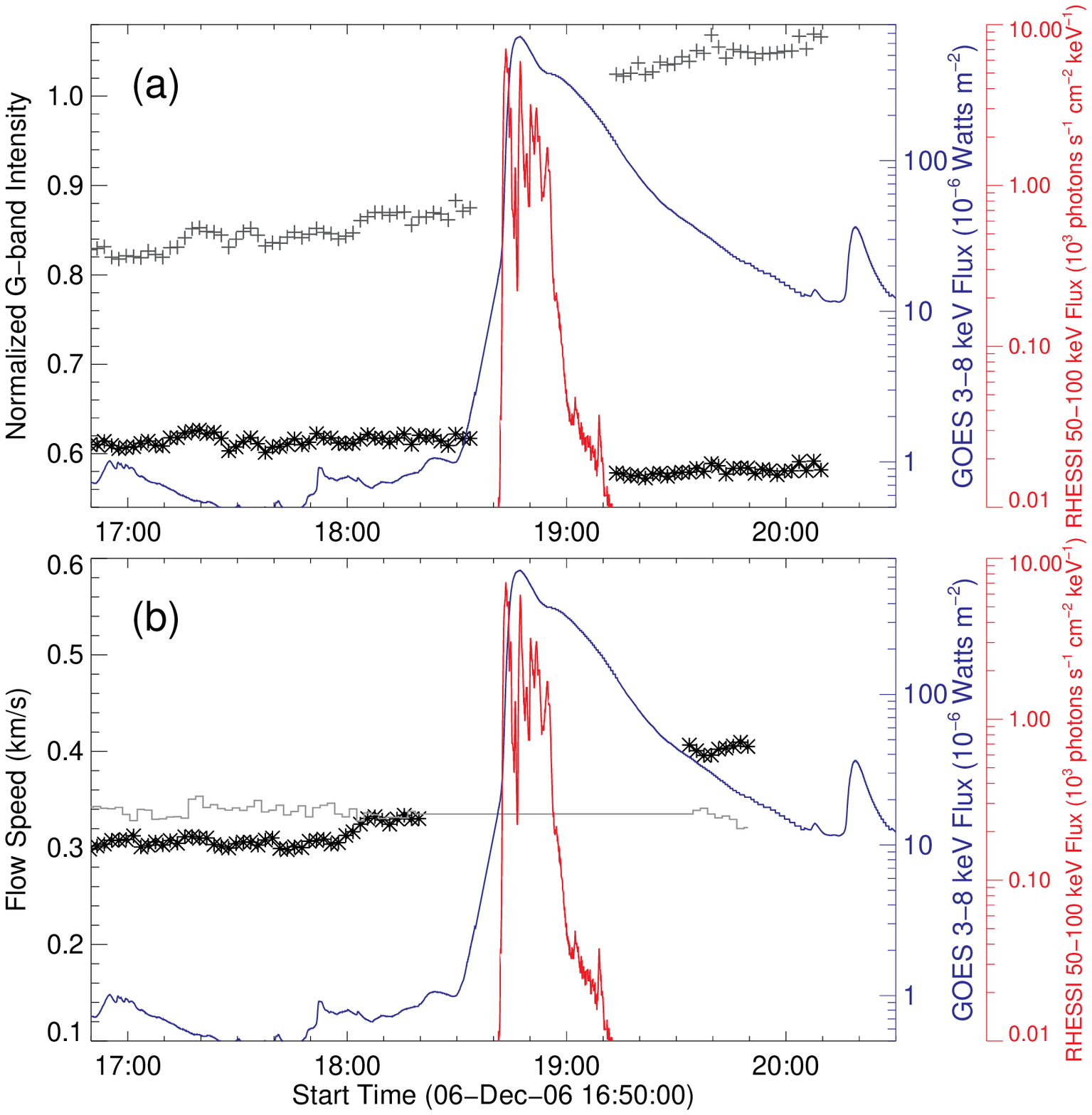}
  \caption{$(a)$: Time profiles of the normalized G-band intensity averaged over the ROI (black data points) and outer decayed penumbral regions (grey plus signs). $(b)$: Time profile of the surface flow speed averaged in the ROI (black data points). The gray curve as a reference shows the temporal evolution of the flow speed averaged in a stable region outlined by the white box in Fig.~\ref{FIG:VELO}. We exclude the data points during the flare to avoid flare transient effects. 10 data points of flow speed at two ends of each time interval are excluded due to the temporal average algorithm.}
  \label{FIG:PLOT}
\end{figure}

This complex $\beta\gamma\delta$ region was located near the east limb (S05E62) during the flare. It is thus necessary to correct the perspective foreshortening effect for accurate measurement of surface flows. The images were remapped to local heliographic coordinates centered at the flaring neutral line\footnote{http://www.csun.edu/$\sim$ndeng/remap/}. Fig.~\ref{FIG:DEPROJ} shows the original and foreshortening-corrected images. The red contour, which is roughly centered at the flaring neutral line, outlines our region of interest (ROI) where the sheared Evershed flow along the neutral line dramatically enhanced right after the X6.5 flare. The magnetic field vector in the sky plane was produced by the High Altitude Observatory (HAO) Milne-Eddington inversion \citep[e.g.,][]{Lites+etal2008ApJ...672.1237L} of the calibrated SP spectra. We then resolved the 180$^\circ$ azimuthal ambiguity of the inverted transverse field using the ``minimum energy'' algorithm \citep{Metcalf1994SoPh..155..235M}. The magnetic field vector was finally transformed from the sky plane to local Cartesian coordinates. Fig.~\ref{FIG:DEPROJ} $(e)$ and $(f)$ show the original B$_{los}$ and the transformed also foreshortening-corrected B$_{z}$ magnetograms, respectively. Some bad data patches in Fig.~\ref{FIG:DEPROJ} $(f)$ are due to either weak signals or failures in the 180$^\circ$ dis-ambiguity calculation for this near limb region.

Local Correlation Tracking (LCT) technique \citep{November+Simon1988ApJ...333..427N} was applied to the well aligned foreshortening-corrected G-band image sequences to acquire horizontal proper motions and their evolution in this active region. A Gaussian shape apodization window with a FWHM of 1200~km \citep{Verma+Denker} was used as the sampling window in the LCT algorithm. To reduce the noise and uncertainty of the individual LCT proper motion map obtained from two successive G-band images, a running temporal average of 10 consecutive LCT proper motion maps traversing 20 minutes was carried out and provides a robust velocity pattern at the center time of each 20-minute interval (see Fig.~\ref{FIG:VELO}). A running temporal average of 20 consecutive LCT proper motion maps was used to evaluate the evolution of the mean horizontal flow speed in the ROI (see Fig.~\ref{FIG:PLOT}(b)).

\section{RESULTS}\label{sec:result}
Fig.~\ref{FIG:IMGS} shows the evolution of NOAA AR 10930 associated with the X6.5 flare in G-band and \CaII\ H images. The flare peaked in white-light and HXR at 18:43:38 UT, showing strong ribbons that separated from the ROI. The sunspot structure remained almost unchanged for at least 4 hours before the flare (see Fig.~\ref{FIG:DEPROJ} $(b)$, Fig.~\ref{FIG:IMGS} $(a)$ and $(b)$). The outer penumbrae, as pointed by the blue arrows, are found dramatically decayed since 19:13~UT (the first available post-flare image) about 20 minutes after the flare ribbon sweeps across these regions. Meanwhile the central penumbra becomes darker. According to the G-band penumbral structure and the \CaII\ H images, it is reasonable to state that there are magnetic connections between P1-P3 and N1-N3 before the flare. After the flare, as pointed out by the pink arrows in Fig.~\ref{FIG:IMGS} $(h)$, some field lines on either side of the neutral line get connected, i.e., new connections between P1-P3 and N4 are formed meanwhile the connections between P1-P3 and N1-N3 would correspondingly be weakened due to this reconnection. The reconnecting magnetic field lines rooted at P1-P3 change from fanning out (connect with N1-N3) that contributes to the outer penumbrae, to inward connection (connect with N4), which naturally results in weaker/stronger horizontal field in the outer/central region relative to the pre-flare configuration and consequently the decay/enhancement of the outer/central penumbrae. Viewing the \CaII\ H movie, we also see the contraction of these newly formed overlying loops throughout the available $\sim$1-hr post-flare phase.

Fig.~\ref{FIG:VELO} depicts the horizontal proper motions in NOAA AR 10930 before and after the flare. The surface flows in the ROI are carried by penumbrae and mainly follow the neutral line direction that is tangential to the umbrae. We therefore call them sheared Evershed flows. Comparing panels $(d)$ $(e)$ with $(a)$ $(b)$, it is quite obvious that the sheared Evershed flows in the ROI dramatically enhanced after the flare. They are more organized along the neutral line with larger speed and extend to larger areas. To explicitly illustrate the flare-associated changes in sunspot structure and surface flows, we further construct the difference maps between post- and pre-flare phases as shown in Fig.~\ref{FIG:DIF}. Similar to previous studies \citep[e.g.,][]{LiuC+etal2005ApJ...622..722L}, the intensity difference map (Fig.~\ref{FIG:DIF} $(a)$) appears bright in the periphery and dark in the central region, manifesting the decay or disappearance of outer penumbrae and darkening of the central structure. The bright/dark features in the flow speed difference map (Fig.~\ref{FIG:DIF} $(b)$) represent surface flow speed increase/decrease. The flow speed clearly increased in most areas of the ROI. In the decayed penumbral regions, some areas show a strong decrease in speed, some others only show a faint decrease in part of areas. The speed decrease mainly occurs at the penumbral edge. The flow azimuth difference map (Fig.~\ref{FIG:DIF} $(c)$) reveals a large change of surface flow direction mainly in the ROI. Flow azimuth changes are also found in some areas of the decayed penumbrae.

Fig.~\ref{FIG:PLOT} illustrates the temporal evolution of the G-band intensity averaged in the ROI and decayed outer penumbrae $(a)$ and the temporal evolution of the horizontal flow speed averaged in the ROI and in a reference region $(b)$. Comparing to the reference curve, the intensity and the flow speed show conspicuous stepwise changes associated with the flare. The rapid changes during the flare occur in as short as 40 minutes or less with change rates much larger than general evolution. The mean horizontal flow speed averaged in the ROI increases from 330$\pm$3.1 to 403$\pm$4.6 \ms, where the values and the errors are the mean and the standard deviation, respectively, of 9 data points in the time profile immediately before and after the flare. The difference of 73 \ms\ is more than 15 times the standard deviation of the data. The changed sunspot structure and the enhanced sheared Evershed flow last for at least 1 hour with the available data indicating a permanent rather than a transient change as a result of the flare. It is worthwhile to point out that there is a noticeable increase of the sheared Evershed flow speed (from 303$\pm$4.0 to 330$\pm$3.1 \ms) at 18:00~UT about 40 minutes before the flare.

We also examined the Michelson Doppler Imager (MDI) 96-min LOS magnetograms and found no increase of the total unsigned magnetic flux in the active region during and after the flare. It implies that no considerable new flux emergence was involved in this flare.

\section{SUMMARY AND DISCUSSION}\label{sec:discussion}

We have presented the rapid stepwise sunspot structure change and enhancement of the sheared Evershed flow along the flaring neutral line associated with the X6.5 flare. These changes are permanent in the sense that they last for at least 1 hour after the flare. Since no signature of new flux emergence was found throughout the event, we exclude the possibility that the enhancement of the sheared Evershed flow may be caused by emerging flux near the neutral line. Considering that the penumbral structure and Evershed flow are strongly controlled by magnetic inclination or horizontal magnetic field, we attribute the observed rapid changes of sunspot structure and surface flow to photospheric magnetic restructuring due to the flare. The central enhanced sheared Evershed flow manifests a more horizontal sheared magnetic field in the photosphere near the flared neutral line. The decayed outer penumbrae indicate the weakening of the horizontal field in the outer region. \CaII\ H images evidence that the originally fanning out field lines at the two sides of the neutral line get connected and contracted over the central region after the flare, which naturally results in overall weakening/strengthening of horizontal magnetic field in the outer/central region and consequently the decay/enhancement of the outer/central penumbrae. Although the highly sheared horizontal fields in the photosphere that give rise to the sheared Evershed flow in the ROI are barely observed in \CaII\ H images, the strengthening and contraction of the newly formed overlying potential fields could be an indicator of the overall trend of the magnetic fields near the neutral line that become more horizontal after the flare. In summary, the observations fit well to the reconnection picture presented in \citet[][Fig.12]{LiuC+etal2005ApJ...622..722L} and are consistent with recent magnetic observations and theoretical predictions of flare-induced photospheric magnetic field change.

Using GONG LOS magnetograms, \citet{Petrie+Sudol2010ApJ...724.1218P} found that this X6.5 flare was associated with the most impressive stepwise magnetic flux changes and Lorentz force budget in their 77 flare samples studied. Moreover, the spatial distribution of their estimated Lorentz force changes \citep[][Fig. 16]{Petrie+Sudol2010ApJ...724.1218P} is co-spatial with the decayed outer penumbral and the enhanced central neutral line regions as illustrated in this letter. Their analysis of the direction of the Lorentz force change does suggest a contraction of the side field lines toward the neutral line resulting in a more horizontal magnetic field in the central region.

Since larger flares tend to produce stronger field configuration changes that eventually result in stronger changes in WL structure and surface flow \citep[e.g.,][]{LiuC+etal2005ApJ...622..722L, ChenW+etal2007ChJAA...7..733C, Petrie+Sudol2010ApJ...724.1218P}, we have not found other smaller Hinode observed events that show enhancement of the central sheared Evershed flow as prominent as this event so far, although some of them do show changes in direction or speed in some areas. For the 2006 December 13 X3.4 flare occurred in the same AR but in the southern part where sheared penumbral flows are also present, \citet{Tan+etal2009ApJ...690.1820T} found, however, no increase of the shear flow after the flare. The X3.4 event is different from the X6.5 event presented in this letter in many aspects. For the X3.4 event, the central sheared penumbrae involve complex ``magnetic channel'' structure \citep{WangH+etal2008ApJ...687..658W}, sunspot rotation, and new flux emergence, which make their structure and flow complicated, thus difficult to interpret. With ground observation, \citet{Deng+etal2006ApJ...644.1278D} detected clues of enhanced shear flow along the flaring neutral line right after the X10 flare on 2003 October 29 in NOAA AR 10486 using limited 18 minutes post-flare data. Different from the sheared Evershed flow presented in this letter that mainly flows in one direction, the sheared penumbral flows in NOAA AR 10486 flow in opposite directions residing at the two sides of the neutral line. The anti-parallel or one-directional sheared penumbral flow depends on whether both umbrae of opposite polarity possess their own penumbrae at the interface where they meet. For NOAA AR 10930 presented in this letter, the left negative polarity umbra is significantly stronger than the right positive ones. The interface is dominated by the penumbrae of the left umbra and the penumbrae of the right umbrae almost vanish at the interface, therefore the sheared Evershed flow of predominant one direction is observed.

Using MDI Dopplergrams for the 2000 July 14 X5.7 Bastille Day flare, \citet{WangH+etal2005ApJ...627.1031W} measured a decrease of Doppler velocity in the outer decayed penumbral segments and an increase in parts of the central darkened penumbral region, suggesting actual weakening/enhancing of the outer/central penumbral structure. The Dopplergrams measure real mass flow but only in the LOS direction. The measured Doppler velocity is subject to the view angle, the location of the object on the Sun, and the flow direction relative to the LOS, therefore Dopplergram alone cannot fully resolve the surface flow. LCT on the other hand performs well for systematic advection motion and has been widely used to estimate solar surface flow. We are aware, though, that LCT measured optical flow may differ from real mass flow because it is based on local contrast that can be affected by other things besides mass flows, such as thermal, oscillation, and wave effects. Nevertheless, the same LCT technique with all the same parameters was applied to the same region, even though the measured velocities may differ from real ones, the trend of the dramatic increase we believe is reliable.

The substantial increase of flow speed in the central region before the flare is intriguing, which may imply a precursor reconfiguration before the eruptive phase of the flare. We do see brightening or formation of some small loops traversing the central neutral line before the flare from the \CaII\ H movie. More events are needed to verify if this pre-flare signature is solid and can be used for flare forecasting.

In summary, this study indicates that besides direct vector magnetograms, high-resolution WL observations of penumbra and coupled Evershed flow can be used as an indirect way to diagnose the magnetic azimuth and inclination as well as their change in the photosphere, which is particularly valuable for flare study even for near limb events.

\acknowledgments
Hinode is a Japanese mission developed and launched by ISAS/JAXA, with NAOJ as domestic partner and NASA and STFC (UK) as international partners. It is operated by these agencies in co-operation with ESA and NSC (Norway). N.D. and D.P.C were supported by NASA grant NNX08AQ32G and NSF grant ATM 05-48952. C.L. and H.W. were supported by NSF grants AGS 08-19662 and AGS 08-49453 and NASA grants NNX 08AQ90G and NNX 08AJ23G. We thank A. Cookson for carefully reading the manuscript.

\end{document}